\documentclass[twocolumn,english,aps,preprintnumbers,amsmath,amssymb,superscriptaddress]{revtex4}
\usepackage[latin9]{inputenc}
\usepackage{amsmath}
\usepackage{graphicx}

\makeatletter
\@ifundefined{textcolor}{}
{%
 \definecolor{BLACK}{gray}{0}
 \definecolor{WHITE}{gray}{1}
 \definecolor{RED}{rgb}{1,0,0}
 \definecolor{GREEN}{rgb}{0,1,0}
 \definecolor{BLUE}{rgb}{0,0,1}
 \definecolor{CYAN}{cmyk}{1,0,0,0}
 \definecolor{MAGENTA}{cmyk}{0,1,0,0}
 \definecolor{YELLOW}{cmyk}{0,0,1,0}
}

\usepackage{dcolumn}% Align table columns on decimal point
\usepackage{bm}% bold math
\usepackage{color}
\usepackage[final]{pdfpages}

\makeatother

\usepackage{babel}

\begin{document}

\preprint{preprint(\today)}

\title{Pressure induced topological quantum phase transition in Weyl semimetal $T_{d}$-MoTe$_{2}$}

\author{Z.~Guguchia}
\email{zurab.guguchia@psi.ch}
\affiliation{Laboratory for Muon Spin Spectroscopy, Paul Scherrer Institute, CH-5232 Villigen PSI, Switzerland}
 
\author{A.M.~dos Santos}
\affiliation{Neutron Scatering Division, Oak Ridge National Laboratory, Oak Ridge, Tennessee 37831, USA}

\author{F.O.~von~Rohr}
\affiliation{Department of Chemistry, University of Zurich, CH-8057 Zurich, Switzerland}
\affiliation{Department of Physics, University of Zurich, CH-8057 Zurich, Switzerland}

\author{J.J. Molaison}
\affiliation{Neutron Scatering Division, Oak Ridge National Laboratory, Oak Ridge, Tennessee 37831, USA}

\author{S.~Banerjee}
\affiliation{Condensed Matter Physics and Materials Science Department, Brookhaven National Laboratory, Upton, NY 11973, USA}

\author{D.~Rhodes}
\affiliation{Department of Materials Science and Engineering, University of Wisconsin-Madison, 1509 University Ave, Madison, WI 53706, USA}

\author{J.-X.~Yin}
\affiliation{Laboratory for Topological Quantum Matter and Spectroscopy, Department of Physics, Princeton University, Princeton, New Jersey 08544, USA}

\author{R.~Khasanov}
\affiliation{Laboratory for Muon Spin Spectroscopy, Paul Scherrer Institute, CH-5232
Villigen PSI, Switzerland}

\author{J.~Hone}
\affiliation{Department of Mechanical Engineering, Columbia University, New York, NY 10027, USA}

\author{Y.J.~Uemura}
\affiliation{Department of Physics, Columbia University, New York, NY 10027, USA}

\author{M.-Z.~Hasan}
\affiliation{Laboratory for Topological Quantum Matter and Spectroscopy, Department of Physics, Princeton University, Princeton, New Jersey 08544, USA}

\author{H.~Luetkens}
\affiliation{Laboratory for Muon Spin Spectroscopy, Paul Scherrer Institute, CH-5232
Villigen PSI, Switzerland}

\author{E.S.~Bozin}
\affiliation{Condensed Matter Physics and Materials Science Division, Brookhaven National Laboratory, Upton, NY 11973, USA}

\author{A.~Amato}
\affiliation{Laboratory for Muon Spin Spectroscopy, Paul Scherrer Institute, CH-5232 Villigen PSI, Switzerland}

%\pacs{74.20.Mn, 74.25.Ha, 74.70.Xa, 76.75.+i}

\begin{abstract}

We report the pressure ($p_{max}$ ${\simeq}$ 1.5 GPa) evolution of the crystal structure of the Weyl semimetal $T_{d}$-MoTe$_{2}$ by means of neutron diffraction experiments. We find that the fundamental non-centrosymmetric structure 
$T_{d}$ is fully suppressed and transforms into a centrosymmertic 1T$^{'}$ structure at a critical pressure of  $p_{\rm cr}$ ${\sim}$ 1.2 GPa. This is strong evidence for a pressure induced quantum phase transition (QPT) between topological to a trivial electronic state. Although the topological QPT has strong effect on magnetoresistance, it is interesting that the superconducting critical temperature $T_{\rm c}$, the superfluid density, and the SC gap all change smoothly and continuously across $p_{\rm cr}$ and no sudden effects are seen concomitantly with the suppression of the $T_{d}$ structure. This implies that the $T_{\rm c}$, and thus the SC pairing strength, is unaffected by the topological QPT. However, the QPT requires the change in the SC gap symmetry from non-trivial $s^{+-}$ to a trivial $s^{++}$ state, which we discuss in this work. Our systematic characterizations of the structure and superconducting properties associated with the topological QPT provide deep insight into the pressure induced phase diagram in this topological quantum material.

\end{abstract}
\maketitle

\section{INTRODUCTION}
 Transition metal dichalcogenides (TMDs) have attracted a lot of attention due to their fascinating physics and
promising functional applications \cite{Soluyanov,SunY,Zheng,WangZ,Weber,KourtisS,KDeng,NXu,Tamai,Kaminski,Xu,Ali1,Balicas1,ZhuZ,PanXC,KangD,QiCava,Bozin}. TMDs share the same formula, MX$_{2}$, where M is a transition metal (for
example, Mo or W) and X is a chalcogenide atom (S, Se and Te). These compounds typically crystallize in a group of related structure types, including 1T$^{'}$- and $T_{d}$-type lattices \cite{Clarke,Puotinen,Zandt,Brown} as shown in Figure 1. The 1T$^{'}$ phase is monoclinic lattice and comprises a distorted octahedral coordination of the metal ions, exhibiting 
pseudo-hexagonal layers with zig-zag edge sharing metal chains. The $T_{d}$ phase is orthorhombic and
both 1T$^{'}$ and $T_{d}$ structures are related by a slight change in the stacking pattern of the layers. Both 1T$^{'}$- and $T_{d}$ are semimetals. The central difference between these two structures is that the 1T$^{'}$ structure exhibits the inversion symmetric space group $P$2$_{1}$/$m$, while the $T_{d}$ phase belongs to the non-centrosymmetric space group ${Pmn}$2$_{1}$.

MoTe$_{2}$ exhibits a 1T$^{'}$-$T_{d}$ structural phase transition, on warming, at $T_{\rm str}$ ${\sim}$ 280 K \cite{QiCava}.  MoTe$_2$, with the non-inversion symmetric orthorhombic $T_{d}$ phase, is a type-II Weyl semimetal \cite{Soluyanov,SunY,WangZ,KourtisS,KDeng,NXu,Kaminski,Tamai,Weber}, where the Weyl Fermions emerge at the boundary between electron and hole pockets. High field quantum oscillation study of the magnetoresistance (MR) for $T_{d}$-MoTe$_{2}$, revealed a nontrivial ${\pi}$ Berry's phase, which is a distinguished feature of surface states \cite{Luo}. Non-saturating magnetoresistance \cite{Xu,Ali1,Balicas1,ZhuZ}, and superconductivity \cite{PanXC,KangD,QiCava}  were also observed in $T_{d}$-MoTe$_{2}$. 
Hence, $T_{d}$-MoTe$_{2}$ represents a rare example of a material with both a topologically non-trivial band structure and superconductivity \cite{Muechler}. The superconducting critical temperature of $T_{d}$-MoTe$_{2}$ is $T_{c}$ ${\simeq}$ 0.1 K \cite{QiCava} at ambient conditions. The application of hydrostatic pressure \cite{QiCava,Guguchia}, the substitution of Te ion by S \cite{Chen} or the creation of Te-vacancies can dramatically enhance $T_{\rm c}$ and lead to a dome-shaped superconductivity in $T_{d}$-MoTe$_2$. Experimental signatures of topological  superconducting order parameter $s^{+-}$ within $T_{d}$  structure were reported by recent muon-spin rotation ${\mu}$SR experiments and $T_{d}$-MoTe$_{2}$ is claimed to be a candidate material for a time reversal invariant topological (Weyl) superconductor (TSC)  \cite{Guguchia,Hosur,Ando,Grushin}. These are special families of materials with unique electronic states, a full pairing gap in the bulk and gapless surface states consisting of Majorana fermions (MFs) \cite{Hosur,Ando,Grushin}. Therefore, the combination of Weyl physics and superconductivity may support Majorana or other exotic surface states in view of their topological nature. These states are of fundamental interest and may eventually be utilised for Quantum computing. For this reason, there is an ongoing effort aimed at achieving superconductivity in such materials and investigating their properties, either with the use of the proximity effect or by other means.  

Besides the enhancement of $T_{\rm c}$, pressure causes the suppression of the non-centrosymmetric orthorhombic $T_{d}$ structure, as it was reported by resistivity experiments \cite{QiCava}. However, resistivity is rather indirect probe for the structural transition and relies only on a weak anomaly, appearing around the structural phase transition $T_{\rm str}$. In order to understand the role of the crystal structure for the occurrence of superconductivity in MoTe$_{2}$, it is essential to directly investigate the structure as a function of pressure using a conventional structural probe such as neutron diffraction. Since pressure has strong effect on superconductivity in MoTe$_{2}$, by investigating structural properties under pressure 
we aim to find the correlation between pressure induced structural and electronic quantum phase transitions in this important TMD system MoTe$_{2}$.

%%%%%%%%%%%%%%%%%%%%%%%%%%%%%%%%%%%%%%%%%%%%%%%%%%%%%%%%%%
\begin{figure*}[b!]
\includegraphics[width=1.0\linewidth]{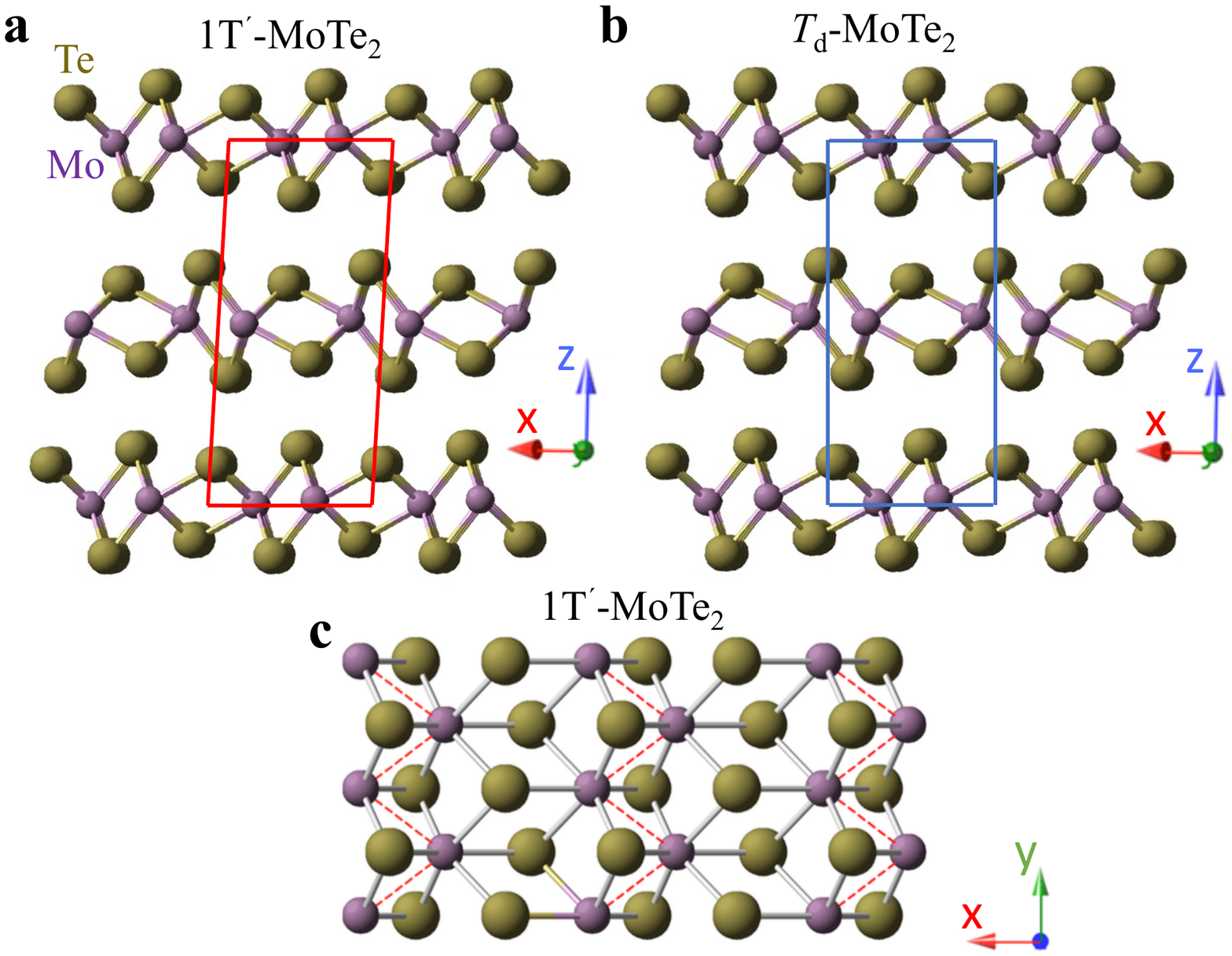}
\vspace{-2.0cm}
\caption{ (Color online) Structural representations (side view) of the centrosymmetric 1T$^{'}$ (a) and non-centrosymmetric $T_{\rm d}$ (b) structures for MoTe$_{2}$. (c) Top view of monolayer constructed from distorted octahedral coordination.}
\label{fig1}
\end{figure*}
%%%%%%%%%%%%%%%%%%%%%%%%%%%%%%%%%%%%%%%%%%%%%%%%%%%%%%%%%%%%%

%%%%%%%%%%%%%%%%%%%%%%%%%%%%%%%%%%%%%%%%%%%%%%%%%%%%%%%%%%%%%%%
\begin{figure*}[t!]
\centering
\includegraphics[width=0.7\linewidth]{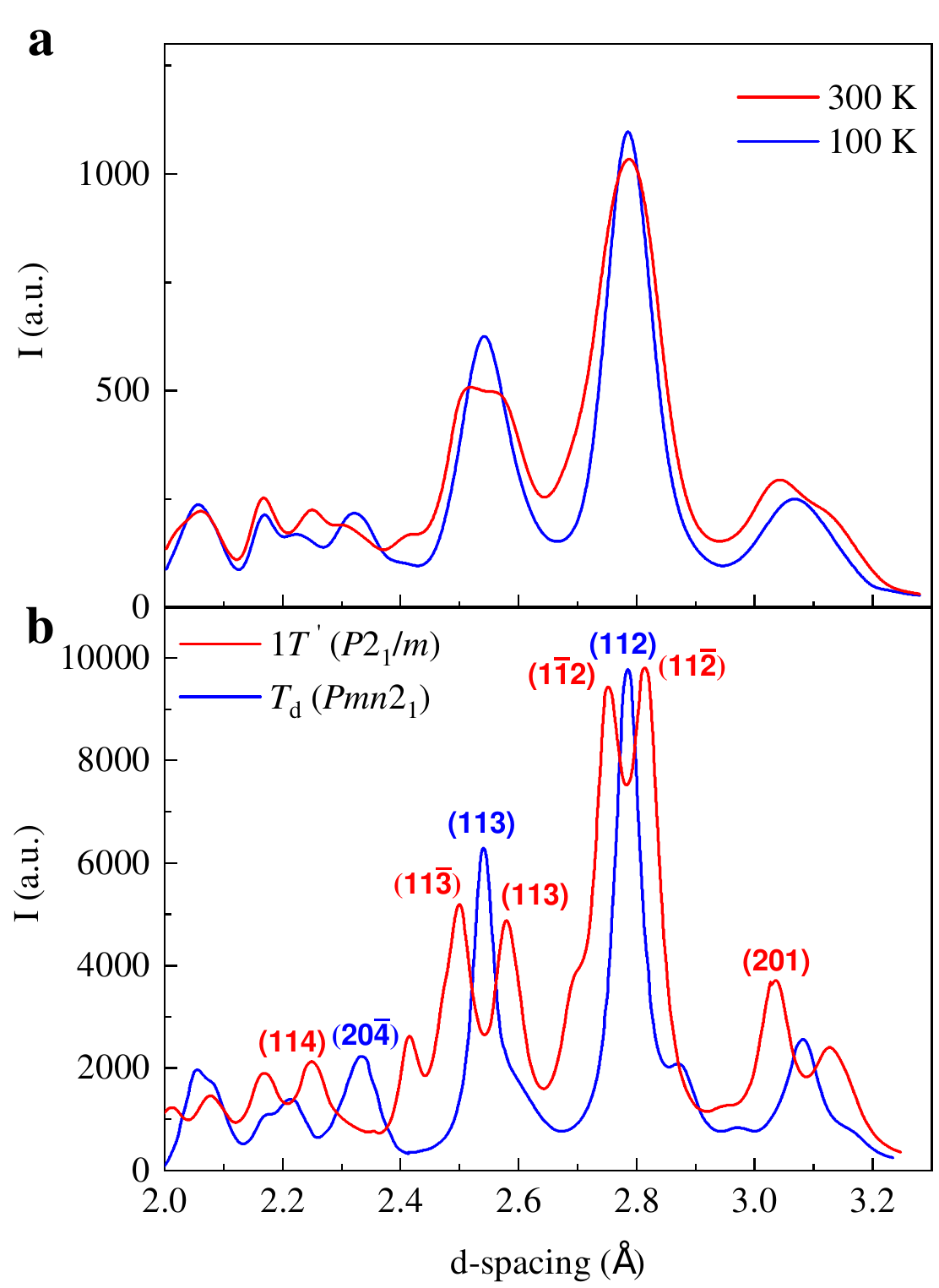}
\vspace{0cm}
\caption{ (Color online) (a) Raw diffraction patterns for 300K and 100K measurements using an incident wavelength of $\lambda=0.18470$~\AA. (b) Bragg profile calculated from unmodified candidate structures. Solid curves in red and blue are the Bragg peaks calculated from the reflections (sharp lines), broadened uniformly with an FWHM=0.1 in $2\theta$\;(=0.059\AA$^{-1}$ in Q). The orthorhombic $(113)$ reflection measured at 100K, splits into the $(11\overline{3})$ and $(113)$ at 300K.}
\label{fig2}
\end{figure*}
%%%%%%%%%%%%%%%%%%%%%%%%%%%%%%%%%%%%%%%%%%%%%%%%%%%%%%%%%%%%%%%

\section{Experimental Details}

High quality single crystals and polycrystalline samples were obtained by mixing of molybdenum foil (99.95 ${\%}$) and tellurium lumps (99.999+${\%}$) in a ratio of 1:20 in a quartz tube and sealed under vacuum. The reagents were heated to 1000$^{\rm o}$C within 10 h. They dwelled at this temperature for 24 h, before they were cooled to 900$^{\rm o}$C within 30 h (polycrystalline sample) or 100 h (single crystals). At 900$^{\rm o}$C the tellurium flux was spinned-off and the samples were quenched in air. The obtained MoTe$_{2}$ samples were annealed at 400$^{\rm o}$C for 12 h to remove any residual tellurium. 

High pressure neutron diffraction experiments on $T_{d}$-MoTe$_{2}$ were carried out at SNAP beamline
of Spallation Neutron Source at Oak Ridge National Laboratory. SNS operates at 60 Hz, and SNAP instrument utilizes time of flight diffraction mode. The center wavelength was set to 2.1 ${\AA}$, corresponding to an incident wavelength spectrum from 0.5 ${\AA}$ to 3.5 ${\AA}$. Two banks of detectors cover +/- 22.5 deg in angular range in and out of plane. Their center was placed at 50 deg scattering angle. This configuration limits the accessible Q-range, but provides increased counting statistics in the region of interest that is most relevant for discriminating between $T_{\rm d}$ and 1T$^{`}$ phases. Pressures up to 1.6 GPa were generated in a single wall piston-cylinder type of cell made of CuBe material, especially designed to perform neutron diffraction experiments under pressure. Daphne oil was used as a pressure transmitting medium. The pressure was always applied at room temperature, followed by cooling the sample to the base temperature (15 K) and carrying out temperature dependent measurements on warming. The pressure was measured by tracking the Raman spectrum of lead powder as a function of pressure. The diffraction experiments were done on the samples from the same batch as the ones, which we previously studied by ${\mu}$SR experiments \cite{Guguchia}.

Total scattering measurements were carried out on the XPD (28-ID-2) beamline at the National Synchrotron Light Source II (NSLS-II), Brookhaven National Laboratory. A finely ground powder of MoTe$_2$ was prepared in an inert Argon chamber, and sealed in 1.02mm (OD) polyimide capillary. Diffraction patterns were collected in a Debye-Scherrer geometry with an X-ray energy of 67.13 keV ($\lambda=0.1847$~\AA) using a large-area 2D PerkinElmer detector (2048$^2$ pixels with 200 $\mu m^{2}$ pixel size). The detector was mounted with a sample-to-detector distance of 345 mm, to achieve a balance between $q$-resolution and $q$-range. The sample was measured at 100K and 300K using an Oxford CS-700
cryostream for temperature control, allowing ample time for the material to thermalize. The experimental geometry, $2\theta$ range, and detector orientation were calibrated by measuring a polycrystalline nickel standard directly prior to data collection, with the experimental geometry parameters refined using the PyFAI program~\cite{kieffer_pyfai_2013}.

\section{Results}

The orthorhombic ($T_{d}$) to monoclinic (1$T^{'}$) structural phase transition in the MoTe$_{2}$ sample at ambient pressure was confirmed by the X-ray total scattering measurements, shown in Fig. 2. In Figure 2a-b, we index the Bragg profile and compare the raw diffraction patterns measured at 100 K and 300 K to powder diffraction patterns calculated from the candidate models using the program VESTA \cite{momma_vesta:_2008}. We plot the data over a $d$-range where the 1$T^{'}$ and $T_{\rm d}$ models have Bragg peaks that are well resolved from one another to highlight distinguishable features that differ between the phases in the measured XRD pattern. The strongest structural signatures of the transition are seen as splitting of the orthorhombic (113) and (112) reflections in monoclinic phase, as it was also reported for WTe$_{2}$ \cite{lu_origin_2016,Tao}.

   Confirming the low temperature $T_{d}$ structure as well as the signature of the structural phase transition in MoTe$_{2}$, we proceed with observations made by neutron scattering measurements under hydrostatic pressure. The most intense peak is (112) at d-spacing of 2.8 ${\AA}$, which was monitored at various temperatures and pressures. The results are shown in Figure 3a-d. Figure 3a shows the temperature evolution of the (112) peak at ambient pressure. 
The peak splits above 285 K, consistent with the $T_{d}$ to 1$T^{'}$ structural phase transition and experimental observations of XRD experiment. The transition temperature $T_{\rm str}$ ${\sim}$ 290 K is higher than the one $T_{\rm str}$ ${\sim}$ 250 K, originally reported from transport experiments \cite{QiCava}. Upon increasing pressure, the temperature, above which the peak splits, decreases, pointing to the suppression of $T_{\rm str}$ with pressure. 
At 1.6 GPa, no single orthorhombic (112) peak is observed within the resolution of our measurement, and monoclinic peaks are observed at all temperature, indicating the full suppression of the orthorhombic $T_{d}$ structure at this pressure. 
Figure 4 shows the pressure dependence of the structural phase transition temperature $T_{\rm str}$, superimposed with the dependence of SC critical temperature (taken from Reference \cite{Guguchia}), measured on the sample from the same batch. This phase diagram shows that the orthorhombic to monoclinic quantum phase transition ($T$=0) takes place at the critical pressure of  $p_{\rm cr}$ ${\sim}$ 1.2 GPa, but $T_{\rm c}$ continues to increase smoothly and persists well beyond the $p_{\rm cr}$. The suppression of the orthorhombic phase as well as the value of the critical pressure is in a very good agreement with very recent single crystal neutron diffraction \cite{Louca}.

\subsection{Discussion}
  
   What is the significance of such a strong negative pressure effect on $T_{d}$ structure and the structural QPT in Weyl semimetal MoTe$_{2}$? Let us start with the fact that $T_{\rm d}$ structure breaks inversion symmetry, while 1$T^{'}$ is inversion symmetric. As low pressure as 1.2 GPa is sufficient to wipe out non-inversion symmetric $T_{\rm d}$ phase. One common point of view in MoTe$_{2}$, mostly supported by calculations, is that Weyl Fermions and their underlying non-trivial band topology, exist only in $T_{\rm d}$ phase. Hence, the structural QPT between $T_{\rm d}$ and 1$T^{'}$ corresponds to the quantum phase transition between topologically non-trivial and a trivial band structure. One of the consequences of such a topological quantum phase transition is  the strong suppression of the magnetoresistance with increasing pressure. For the pressure values $p$ ${\textgreater}$ $p_{\rm cr}$, MR is fully suppressed \cite{ParkLee}. This was explained by the fact that MR is very sensitive to a hole-to-electron concentration ratio $N_{\rm h}$/$N_{\rm e}$ and by driving the system into 1$T^{'}$ phase, the strong deviation from the optimal ratio takes place, leading to the suppression of MR. In fact, MR is strongly suppressed already well below $p_{\rm cr}$, which is ascribed to the presence of a significant band tilting around the Weyl nodes under pressure within the $T_{\rm d}$ phase, as shown by calculations \cite{Louca}. When the Weyl nodes fully disappear at $p_{\rm cr}$, than this leads to the full suppression of MR \cite{ParkLee}. 

Although the topological QPT has a strong effect on MR \cite{ParkLee}, it is interesting that the superconducting critical temperature $T_{\rm c}$ as well as the SC gap ${\Delta}$ (Figure 4, only the large SC gap \cite{Guguchia} is shown) changes smoothly across $p_{\rm cr}$ and no sudden effects are seen due to the suppression of $T_{d}$ structure. This implies that the $T_{\rm c}$ and thus the SC pairing strength is unaffected by the topological QPT. This can be understood in light of Hall conductivity measurements, showing that a pair of electron and hole bands are dominant in the $T_{d}$ structure and the carrier densities on these bands increase smoothly with pressure and no sudden changes are observed across $p_{\rm cr}$. Having this in mind as well as our previous work \cite{Guguchia}, according to which $T_{\rm c}$ scales with the superconducting carrier density, one can understand the insensitivity of $T_{\rm c}$ to topological QPT.  

Since there is only a subtle change in the Fermi surface (FS) topology \cite{ParkLee} at $p_{\rm cr}$, bulk SC quantities such as $T_{\rm c}$, ${\Delta}$ and the superfluid density, are not affected by QPT. However, one more physical parameter that we should consider is the symmetry of the SC order parameter. Previously, we showed that the superconductor $T_{d}$-MoTe$_{2}$ represents a time-reversal-invariant Weyl semimetal, which has broken inversion symmetry below $p_{\rm cr}$. Moreover, it is a two-gap superconductor both below and above $p_{\rm cr}$ (we have one point just above $p_{\rm cr}$) and the gaps are momentum independent on each Fermi surface. Following the theoretical studies of time-reversal-invariant topological superconductivity in Weyl semimetals, the nontrivial Berry curvature at the FSs of WSMs allows the TSC to be realized for a pairing function with no special momentum dependence. A simple formula \cite{Hosur,Ando,Grushin} relates the FS Chern number to the topological invariant ${\nu}$ for a time-invariant TSC, considering a set of FSs with Chern numbers {$C_{\rm j}$}:

\begin{equation}
\begin{aligned}
{\nu} = \frac{1}{2} {\sum_{j \in FS}}C_{\rm j}{sgn}({\Delta}_{j}), 
\label{eq3}
\end{aligned}
\end{equation}  

where $\Delta_{\rm j}$ is the pairing gap function on the $j$-th FS.  A TSC is implied by ${\nu} {\neq}$  0. Therefore, as previously discussed \cite{Hosur}, a necessary condition for TSC is that ${\Delta}_{\rm 1}$ and ${\Delta}_{\rm 2}$ have opposite signs. Thus, $T_{d}$-MoTe$_{2}$ is a natural candidate for topological $s^{+-}$ SC order parameter within non-centrosymmetric $T_{d}$ structure since the nontrivial Chern number is already provided by the band structure. The $s^{+-}$ order parameter in $T_{d}$-MoTe$_{2}$ is also supported by strong disorder dependence of $T_{\rm c}$ \cite{Guguchia,Balicas1}. It was shown that the violation of inversion symmetry stabilises the state where the ferromagnetic exchange coupling is greater than or comparable to the repulsive density-density interactions, which than results in topological superconducting state \cite{Hosur}. Once the inversion symmetry is recovered than the TSC is quenched. This means that at pressure above $p_{\rm cr}$, where  crystal structure has inversion symmetry, the trivial $s^{++}$ state (where ${\Delta}_{\rm 1}$ and ${\Delta}_{\rm 2}$  have the same sign) should be favoured over the topological $s^{+-}$ state. While a direct proof for such a pressure induced QPT between between $s^{+-}$ and $s^{++}$ states in MoTe$_{2}$ is clearly lacking, we conjecture this based on solid experimental evidence gleaned from the quantum phase transition between non-centrosymmetric to a centrosymmertic  structure and our previous observation of two-gap superconductivity. This paves the way for rigorous experimental as well as theoretical explorations of this idea. One immediately useful study would be that of the effects of impurities on superconductivity above
$p_{\rm cr}$. This should discriminate/differentiate between topological $s^{+-}$ and ordinary $s^{++}$ state, since according to the theoretical proposal TSC is much more sensitive to disorder than the ordinary $s^{++}$ superconductivity.

%%%%%%%%%%%%%%%%%
\begin{figure*}[t!]
\includegraphics[width=1.1\linewidth]{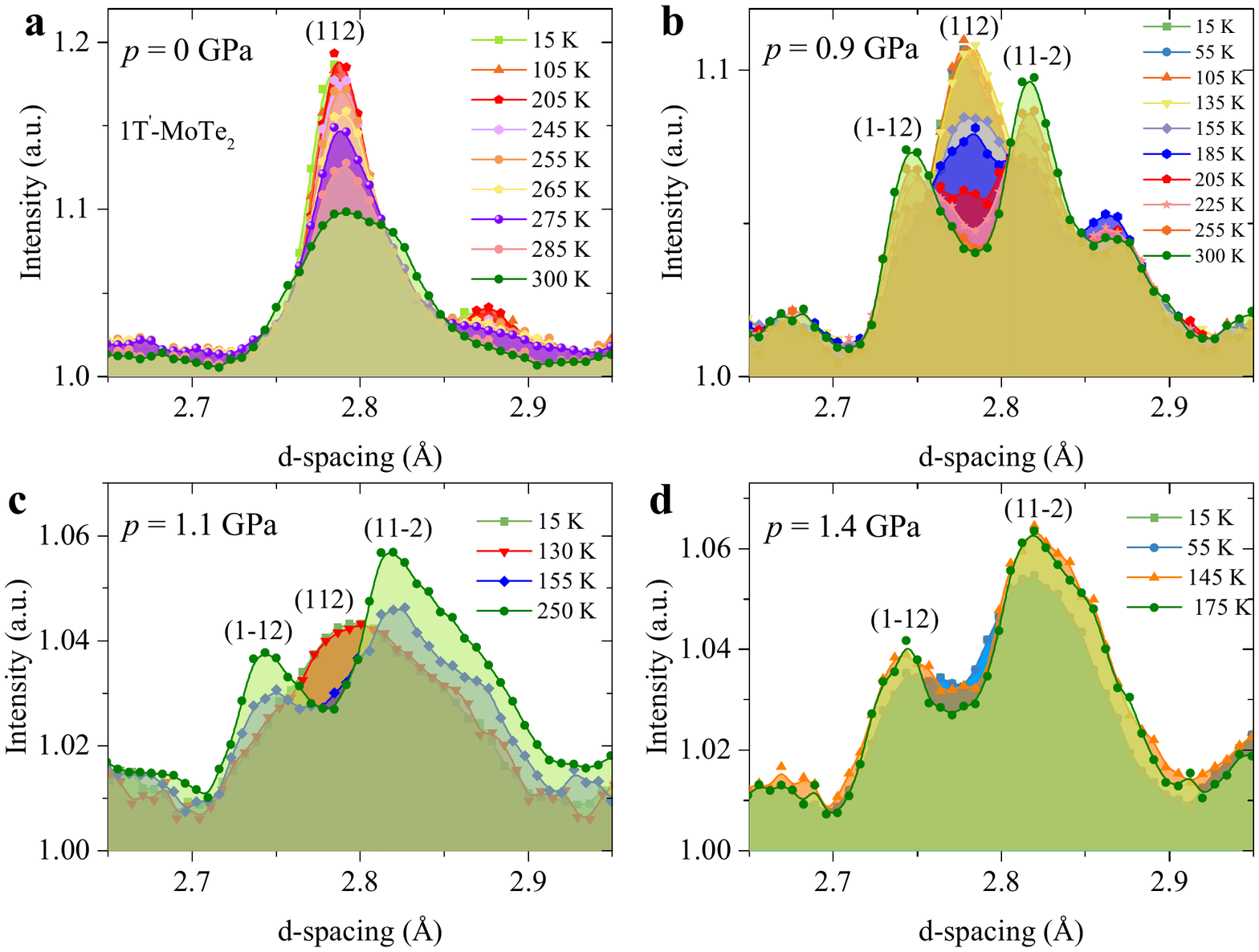}
\vspace{-2.2cm}
\caption{(Color online) The temperature dependence of (112) Bragg peak of MoTe$_{2}$, recorded at various pressures: (a) $p$ = 0 GPa, (b) $p$ = 0.9 GPa, (c) $p$ = 1.1 GPa, and (d) $p$ = 1.4 GPa.}
\label{fig3}
\end{figure*}
%%%%%%%%%%%%%%%  
 
 %%%%%%%%%%%%%%%%%%%%%%%%%%%%%%%%%%%%%%%%%%%%%%%%%%%%%%%%%%%%%%%
\begin{figure*}[t!]
\centering
\includegraphics[width=0.7\linewidth]{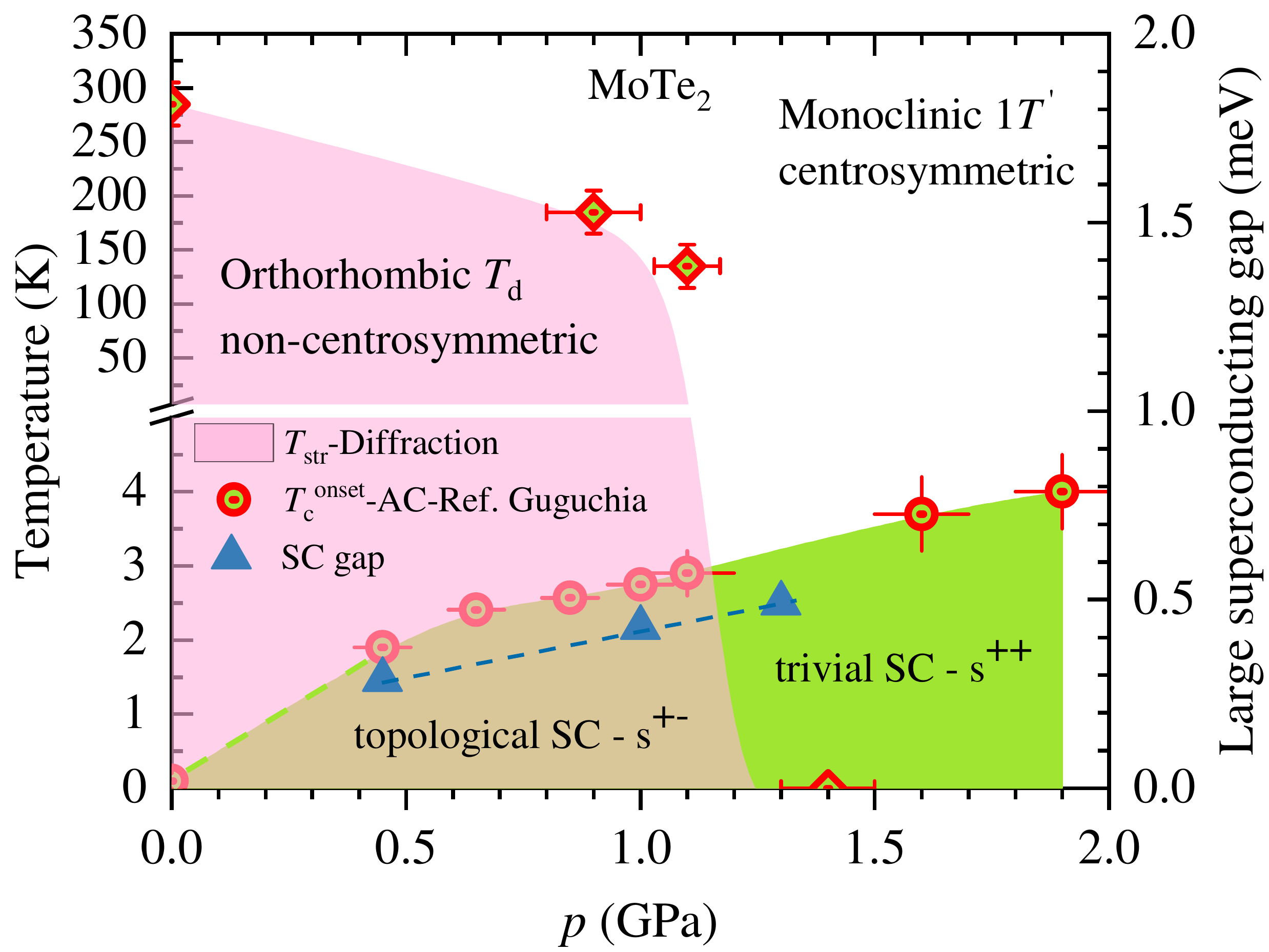}
\vspace{0cm}
\caption{ (Color online)  (a) The pressure dependence of the superconducting critical temperature $T_{\rm c}$ (the red circles), the large SC gap ${\Delta}_2$ (the blue triangles) and the monoclinic to orthorhombic structural phase transition temperature $T_{\rm str}$ (the red diamonds). The $T_{\rm c}$ and the ${\Delta}_2$ values are taken from Ref. 25.}
\label{fig4}
\end{figure*}
%%%%%%%%%%%%%%%%%  

\section{CONCLUSIONS}

In conclusion, we studied the hydrostatic pressure evolution of a monoclinic 1T$^{'}$ to orthorhombic $T_{d}$ structural phase transition temperature $T_{\rm str}$ using neutron diffraction. The quantum phase transition from 
non-centrosymmetric $T_{d}$ to a centrosymmetric 1T$^{'}$ structure is observed under pressure with the critical pressure of $p_{\rm cr}$ ${\sim}$ 1.2 GPa. No obvious impact of such a QPT on SC critical temperature, the SC gap(s) and the superfluid density is observed, suggesting that only subtle changes take place in the Fermi surface across $p_{\rm cr}$
despite the transition from a topological to a topological-trivial band structure. However, due to the recovery of inversion symmetry above $p_{\rm cr}$, the trivial $s^{++}$ superconducting state should be favoured over the topological $s^{+-}$ state. Hence, this system may likely be the first known example of a pressure induced change between topological ($p$ ${\textless}$ $p_{\rm cr}$) and trivial ($p$ ${\textgreater}$ $p_{\rm cr}$) SC states.

%%%%%%%%%%%%%%%%%
\section{Acknowledgments}~
A portion of this research used resources at the Spallation Neutron Source, a DOE Office of Science User Facility operated by the Oak Ridge National Laboratory. Work at Brookhaven National Laboratory was supported by US DOE, Office of Science, Office of Basic Energy Sciences under contract DE-SC0012704. Research at Columbia was supported by US NSF DMR-1610633 and the Reimei Project of the Japan
Atomic Energy Agency. Z. Guguchia gratefully acknowledges the financial support by the Swiss National Science Foundation (SNF fellowship P300P2-177832). The work at the University of Zurich was supported by the Swiss National Science Foundation under Grant No. PZ00P2-174015.

\section{Competing interests} The authors declare that they have no competing interests.\\

\section{Data and materials availability} All data supporting the stated conclusions of the manuscript are in the paper. Additional data are available from the authors upon request.\\

%\includepdf[pages={-}]{MoTe2Supplementary.pdf}
%\includepdf[pages={3}]{MoTe2Supplementary.pdf}
%\includepdf[pages={4}]{MoTe2Supplementary.pdf}

\end{document}